# THE DISTRIBUTION OF COSMOLOGICAL $\gamma$-RAY BURSTS

Ehud Cohen and Tsvi Piran

*Racah Institute for Physics, The Hebrew University, Jerusalem, Israel 91904*

## Abstract

We compare the burst distribution of the new (2B) BATSE catalogue to a cosmological distribution. The observed distribution agrees well with a cosmological one, however, it is insensitive to cosmological parameters such as $\Omega$ and $\Lambda$. The bursts are not necessarily standard candles and their luminosity can vary by up to a factor of ten. The maximal red shift, $z_{max}$, of bursts longer than 2 sec is $2.1^{+1}_{-.7}$ (assuming no evolution). The present data is insufficient to determine maximal red shift, $z_{max}$, of bursts shorter than 2 sec.

*Subject headings:* gamma-rays: bursts - Cosmology: Observations

## 1 Introduction

The recent observations of the BATSE experiment on the COMPTON GAMMA RAY OBSERVATORY have suggested that $\gamma$-ray bursts (GRBs) originate from cosmological sources (Meegan, 1992) (see however, e.g. Podsiadlowski, Rees & Ruderman, 1994 for a different point of view). The count distribution, $N(C)$, in the BATSE 1B catalogue agrees well with a cosmological distribution (Piran, 1992; Mao & Paczyński, 1992; Dermer, 1992; Schmidt, 1992; Wickramasinghe *et al.* 1993, Mao, Narayan & Piran, 1994, denoted MNP). Additionally there is evidence (Norris, *et al.* 1994,1995, Nemiroff, *et al.* 1994), for the predicted (Piran, 1992, Paczyński, 1992) correlations between the duration, the strength and the hardness of the bursts (see however also Mitrofanov *et al.* 1994 and Fenimore 1994). With the release of the new (2B) BATSE catalogue (Meegan, *et al.* 1994) we analyze here the count distribution in terms of a cosmological model.

We calculate the likelihood function to obtain the observed bursts' counts distribution, $N(C)$, from a variety of theoretical cosmological models. We show, first, that the observed count distribution in the 2B catalogue agrees with a cosmological distribution. The observed distribution is, however, insufficient to distinguish between different cosmological models. We demonstrate that the bursts are not necessarily standard candles and that their luminosity can vary by up to a factor of ten. We conclude by comparing our analysis with the results of Norris *et al.* (1994,1995) on the detection of cosmological time dilation and with the analysis of Fenimore *et al.* (1993) of the combined 1B data and PVO data.

## 2  Basic Model

We review here, briefly, the basic features of a cosmological source count distribution (see e.g. Weinberg, (1973) for a general population of sources and e.g. Piran, (1992) for GRBs). We consider a detector with a fixed energy range, $\Delta E$, that operates for a fixed time span, $\Delta t$, and sources with a count spectrum: $N(E) \propto E^{-\alpha}$. The count rate at the detector from a source at a redshift $z$ is:

$$C(\tilde{L}, z) = \frac{\tilde{L}(1+z)^{2-\alpha}}{4\pi d_l^2(z)} \quad (1)$$

where $\tilde{L}$ depends on the luminosity of the source, $L$, in the relevant energy range, $\Delta E$, on the average energy $\bar{E}$ and on the observation time, $\Delta t$: $\tilde{L} = L(\Delta E)\Delta t/\bar{E}$. The luminosity distance is $d_l(z)$. The number, $N(>C)$, of events with an observed count rate larger than $C$ is:

$$N(>C) = 4\pi \int_0^\infty n(L) dL \int_0^{z(C,L)} \frac{d_l^2}{(1+z)^3} n(z) \frac{dr_p(z)}{dz} dz \quad (2)$$

where $r_p(z)$ is the proper distance to a redshift $z$ and $z(C, L)$ is obtained by inverting Eq. 1. The luminosity function is $n(L)$ and the intrinsic rate (the number of events per unit proper volume and unit proper time at redshift, $z$) is $n(L)$. The distribution, $N(C)$, depends on the cosmological parameters: $\Omega$ and $\Lambda$, and on the source parameters: $\alpha$, $n(z)$ and $n(L)$.

We assume that the sources have a spectral index $\alpha = 1.5$ (Schaefer *et al.* ; 1992). The resulting distribution is sensitive to the value of $\alpha$ and therefore we quote the parameters obtained for $\alpha = 2$. The intrinsic evolution, $n(z)$ and the luminosity function $n(L)$ are unknown. We have chosen $n(z)$ of the form:

$$n(z) = n_0(1+z)^{-\beta}, \quad (3)$$

Note that Piran (1992) uses $n(t) = n_0 t^{\beta'}$ and for small $z$ $\beta' \approx -3\beta/2$.

A standard candle distribution,

$$n(L) = n_0 \delta(L - L_0), \quad (4)$$

is characterized by one parameter, $z_{max}$, the maximal $z$ at the source is detected (obtained by inverting Eq. 1 for $C = C_{min}$ and $L = L_0$). The parameter $n_0$ is determined from the analysis. We consider more general distributions that depend on two parameters. We examine a decreasing power law luminosity function with a minimal luminosity $L_0$:

$$n(L) = [n_0(s-1)/L_0](L/L_0)^{-s}, \quad \text{for} \quad L > L_0 \quad (5)$$

and an increasing power law with a maximal luminosity. These distributions are characterized by $z_{max}(\langle L \rangle)$ and by the power law index $s$. We also examine a maximally non - smooth distribution, a two standard candles distribution:

$$n(L) = n_1 \delta(L - L_1) + n_2 \delta(L - L_2), \qquad (6)$$

where the ratio $n_1/n_2$ is chosen so that the number of bursts observed from each group is equal. This distribution is characterized by $z_{max}(\langle L \rangle)$ and by the luminosity ratio, $L_1/L_2$.

## 3 The Data

The 2B catalog available in the public domain contains a list of all gamma-ray bursts that triggered the BATSE detectors up to 1993 March 9. The catalog contains the ratio $C_{max}/C_{min}$ for 412 bursts, where $C_{max}$ is the maximum count rate and $C_{min}$ is the detection threshold and a duration table for 434 bursts (Meegan et al. 1994). BATSE's background varies as a function of time, so that the detection threshold $C_{min}$ does not remain constant. To obtain a regular sample we select a constant threshold $C_{cut}$ ($C_{cut}(1024ms) = 287$ and $C_{cut}(64ms) = 72$), and prune the data to contain only bursts which satisfy $C_{min} \leq C_{cut} \leq C_{max}$ (Petrosian, 1993; MNP). Following MNP we ignore the 256ms channel which seems to be more prone to irregularities.

Kouveliotou *et al.* (1993) have shown that GRBs are bimodal in duration with short ($\delta t_{90} \leq 2$ s) and long ($\delta t_{90} \geq 2$ s) bursts. Since BATSE is more sensitive to the long bursts (MNP) it is meaningless to combine both sub-classes to one distribution and we have to separate the BATSE population to two sub-groups. In order not to exclude bursts which do not have a duration record we have used the relation (MNP):

$$\frac{(C_{max}/C_{cut})_{1024}}{(C_{max}/C_{cut})_{64}} \begin{cases} > 1, & \text{for long bursts,} \\ < 1, & \text{for short bursts,} \end{cases} \qquad (7)$$

to add 15 bursts to our set. After excluding bursts which do not satisfy with all the previous requirements we have 238 long bursts, and 50 short bursts.

In principle we would like to compare both the short and the long sub-populations to theoretical cosmological distributions. However, some of the short bursts are shorter than 64ms. The count rate in this channel corresponds for these bursts to the total fluence of the burst and it underestimates the peak intensity of these events. This will distort the count distribution $N(C)$. We exclude, therefore, the short bursts from this analysis and compare here only the long sub-class with the theoretical models. As far as we know all previous studies of the BATSE data ignored this problem and included the short bursts in their analysis. This study is the only one to date which proceeds to accomplish an accurate cosmological analysis and interpretation for LONG bursts only, using the 2B catalog.

## 4 Results

The theoretical models depend on five unknown parameters; $\Omega, \Lambda, z_{max}, s$ and $\beta$ as well as on $\alpha$. We calculate the likelihood function (see e.g. Press *et al.* , 1993) over this five dimensional parameter space and find the range of acceptable models (those whose likelihood function is not less than 1% of the maximal likelihood). We then proceed to perform a KS (Kolmogorov-Smirnov) test (see e.g. Press *et al.* 1993) to check whether the model is acceptable. We present in the following several two dimensional cuts through the parameter space.

We begin with $\Omega = 1$ and $\Lambda = 0$ and standard candles with no source evolution. For this simple model the likelihood function peaks at $z_{max} = 2.1$ and the allowed range at a 1% confidence level is: $1.4 < z_{max} < 3.1$ (for a spectral index $\alpha = 1.5$). For a spectral index $\alpha = 2$ the 1% confidence range is: $z_{max}^{(\alpha=2)} = 1.5^{+.7}_{-.4}$. Figure 1 depicts the observed number count distribution for the long bursts (using the 1024ms channel) and the theoretical counts distributions for the allowed range of $z_{max}$.

Figure 2 depicts the contour lines of the likelihood function for the long bursts distribution in the ($\Omega,z_{max}$) plane. The likelihood function is practically independent of $\Omega$ in the range: $0.1 < \Omega < 1$. The likelihood function is also insensitive (figure not shown) to the cosmological constant $\Lambda$ (in the range $0 < \Lambda < 0.9$, in units of the critical density). Following these results we are left only with the intrinsic parameters of the bursts distribution and we present in the rest of the paper results for cosmologies with $\Omega = 1$ and $\Lambda = 0$.

Figure 3 describes the likelihood function in the ($z_{max}, \beta$) plane for sources with varying intrinsic rate. The banana shaped contour lines show that a distribution with no intrinsic $z$ dependence ($\beta = 0$) is equivalent to one with an increasing number of bursts with cosmological time ($\beta > 0$) with a lower $z_{max}$. This tendency saturates at high intrinsic evolution (large $\beta$), for which the limiting $z_{max}$ does not go below $\approx .5$ and at very high $z_{max}$, for which the limiting $\beta$ does not decrease below -1.5.

We examined three luminosity functions: a decreasing power-law with a minimal luminosity, Eq. 5, an increasing power-law with a maximal luminosity and a double standard candles distribution, Eq. 6. Figure 4 depicts the likelihood function in the $(s, z_{max}(\langle L \rangle))$ plane for luminosity functions of the form Eq. 5. We find that the likelihood function is quite insensitive to $s$ as long as $s \gtrsim 2.2$ (note that $\langle L \rangle$ diverges as $s \rightarrow 2$). The average luminosity with $s = 2.2$ is $5L_0$, hence the luminosity varies, effectively by a factor of ten in this distribution. The likelihood function doesn't go below 70% of the maximal value for the whole range of power indices, $s$, for an increasing power law with an upper cutoff. This distribution is dominated by the strongest sources and the addition of weaker sources hardly effects $N(C)$. When we considered two standard candles sources, Eq. 6, the 10% likelihood is around a luminosity ratio of 10 between the sources. The KS test gives a probability of 80% for a luminosity ratio of 14. These results demonstrate that the BATSE data alone do not require a narrow luminosity function.

PVO has been detecting GRBs for about 14 years - but with a lower sensitivity than BATSE. The PVO count distribution, $N(C) \propto C^{-3/2}$, corresponds to a homogeneous distribution indicating that PVO is detecting only nearby GRBs whose distribution is not influenced by cosmological effects. In principle one should combine the BATSE and the PVO data. However, the PVO data are not available publicly at this stage. We can still incorporate the main PVO result in our analysis using synthetic data imitating the PVO data: a data set with the same number of bursts as observed by PVO, with an Eucleadian count distribution, $N(C) \propto C^{-3/2}$, and a sensitivity limit larger by a factor of 20 than BATSE (Fenimore *et al.* (1993). We perform a maximum likelihood analysis (see Figure 3) for the combined data and we find that the combined distribution has $z_{max} \approx 0.87$ (for standard candles with no source evolution). The allowed $z_{max}$ range at 1% confidence is $0.65 < z_{max}(PVO + BATSE) < 1.2$. This reproduces the main conclusion of Fenimore *et al.* (1993) who combined the full (long and short) 1B data with the PVO data. It is not known what is the fraction of long vs short PVO bursts. If we assume that the ratio of short to long bursts in PVO is the same as in BATSE, these results are slightly modified and we obtain: $0.55 < \tilde{z}_{max}(PVO + BATSE) < 1.05$.

When we add luminosity functions (of the same kind described earlier) we find that the combined data require a narrow luminosity function. For example when we consider the two delta function distribution, the allowed luminosity ratio (at 1%) confidence is 2.5 compared to a factor of 10 in the BATSE data alone. Source evolution has the same effect as seen in the BATSE data - $z_{max}$ decreases when we consider sources whose intrinsic burst rate increases with cosmological time (see figure 3).

The result that $z_{max} = 0.87$ is the maximal red shift of the BATSE data should be compatible with the BATSE data on its own. However, a KS test rules out $z_{max} = 0.87$ for the BATSE 2B data: The probability for $z_{max} = 0.87$ of 2B data alone is less than $2 \cdot 10^{-6}$ (for $\alpha = 1.5$). The 1% upper limit of PVO+BATSE ($z_{max}(PVO + BATSE) = 1.2$) is ruled out at a $10^{-3}$ level. The discrepancy between the two data sets improves slightly when we include intrinsic source evolution (see Figure 3) but even with $\beta = 2.5$ the overlap between the likelihood functions of the two data sets is at a significance level of $\approx 0.3\%$. A possible alternative resolution is that this discrepancy arises from systematic errors in matching the sensitivity and the triggering algorithms of the two detectors.

## 5  Conclusions

For standard candles with no source evolution the maximal red shift up to which the long bursts are observed is $z_{max} = 2.1^{+1.1}_{-0.7}$. With an estimated BATSE detection efficiency of $\approx 0.3$, this corresponds to $2.3^{-0.7}_{+1.1} \cdot 10^{-6}$ events per galaxy per year (for a galaxy density of $10^{-2} h^3$ Mpc$^{-3}$; Kirshner *et. al*, 1983). The rate per galaxy is independent of $H_0$ and is only weakly dependent on $\Omega$. For $\Omega = 1$ and $\Lambda = 0$ the typical energy of a burst whose observed fluence is $F_7$ (in units of $10^{-7}$ergs/cm$^2$) is $7^{+11}_{-4} \cdot 10^{50} F_7$ ergs. These numbers vary slightly if the bursts have a wide luminosity function. The distance to the sources decreases and correspondingly the rate increases and the energy decreases if the spectral index is 2 and not 1.5. The rate increases and the luminosity decreases if there is a positive evolution of the rate of bursts with cosmological time.

Several luminosity functions with effective width of up to factor of ten fit the data. This is slightly wider than the allowed width (approximately a factor of six) obtained by Horack and Emslie (1994), who use a different statistical approach ("moment analysis") and whose data base is the BATSE 1B catalogue. The two results are consistent in view of the different techniques, different data bases and different models for the luminosity functions used. This width is comparable to, and even wider than, the width of some observed luminosity distributions such as the luminosity functions of different types of supernovae.

In agreement with Fenimore *et al.* (1993) we find that the combined BATSE data and the simulated PVO data corresponds to a cosmological distribution with $z_{max} \approx 0.87$. The combined data also require a narrow luminosity function. This result is, however, inconsistent with the 2B data on its own. A luminosity function or source evolution do not change this conclusion. This inconsistency suggests that there are systematic errors in matching the PVO and BATSE data. It might be rectified if either the sensitivity of PVO relative to BATSE has been overestimated, or if the PVO detection efficiency relative to BATSE is lower than expected.

Norris *et al.* (1994,1995) found that the dimmest bursts are longer by a factor of $\approx 2.3$ compared to the bright ones (note however, that Mitrofanov *et al.* (1994) find no evidence for time dilation). With our canonical value of $z_{max} = 2.1$ the bright bursts originate at $z_{bright} \approx 0.2$. The corresponding expected ratio due to cosmological time dilation, 2.6, is in agreement with this measurement. If true, this results will support the cosmological model and imply that (i) intrinsic source evolution is insignificant and (ii) the low $z_{max}$ obtained form the combined PVO and BATSE data is rules out. Fenimore & Bloom (1994) find, on the other hand, that when other effects are included, this time dilation corresponds to $z_{max} > 6$. In this case (ii) is still valid. However, (i) is invalid since $z_{max} > 6$ requires a significant negative intrinsic evolution: $\beta \approx -1.5 \pm 0.3$.

The GRB count distribution is compatible with a cosmological one. Unlike preliminary hopes, (Trimble, 1994) we find that it is impossible to distinguish between different cosmological models using the available GRB data. Firstly, $N(C)$ is rather insensitive to $\Omega$ and $\Lambda$. Additionally, like in other cosmological sources cosmological evolution can easily mask some cosmological effects. However, not all hope is lost. If GRBs result from neutron star mergers (Eichler *et al.* 1989), than detectors like LIGO and VIRGO could detect gravitational radiation signals from those events in coincidence with GRBs (Piran, 1992; Kochaneck & Piran, 1993). This would enable us to measure the distance to those objects (Schutz, 1986). This additional information, which is not available to any other cosmological objects, might enable us to determine cosmological parameters using the GRB distribution after all.

We thank Tsafrir Kolatt, Jay Norris and Ramesh Narayan for helpful discussions and comments. This research was supported by a BRF grant to the Hebrew University and by a NASA grant NAG5-1904.

---

Figure 1: The normalized observed number counts distribution of the long bursts (with the 1024msec channel) and the theoretical cosmological distribution. The theoretical distribution is shown for three values of $z_{max}$: 2.1 (dotted line) 1.4 (long dashed line) and 3.1 (short dashed line) which correspond to the maximal likelihood value and the two 1% bounds. Also shown is a theoretical curve obtained for $z_{max} = 0.87$ (dash-dot line: the best fit for the joint PVO and BATSE data). The descrepancy between this curve and the BATSE data is apparant. The curves are calculated for $\Omega = 1$, $\Lambda = 0$, $\alpha = -1.5$ and for standard candles with source evolution.

Figure 2: Contour lines of likelihood function (levels 33%, 10%, 3.3% 1% etc.. of the maximum, marked by +) in the $(\Omega, z_{max})$ plane for standard candles (Eq. 4), with no source evolution ($\beta = 0$) and $\alpha = 1.5$. The cosmological constant $\Lambda = 0$.

Figure 3: Gray scale map of the likelihood function (levels 33%, 10%, 3.3% 1% etc.. of the maximum) in the $(\beta, z_{max})$ plane for standard candles sources (Eq. 4) with $\alpha = 1.5$ and "standard cosmology" $\Omega = 1$, $\Lambda = 0$ for intrinsic evolution given by Eq. 3. Super imposed on these map is the similar likelihood function for the combined BATSE and PVO data

Figure 4: Contour lines of likelihood function (levels 33%, 10%, 3.3% 1% etc.. of the maximum, marked by +) in the $(s, z_{max}(\langle L \rangle))$ plane for "standard cosmology" $\Omega = 1$, $\Lambda = 0$ and source with no intrinsic evolution ($\beta = 0$) and $\alpha = 1.5$ for a luminosity function given by Eq. 5.

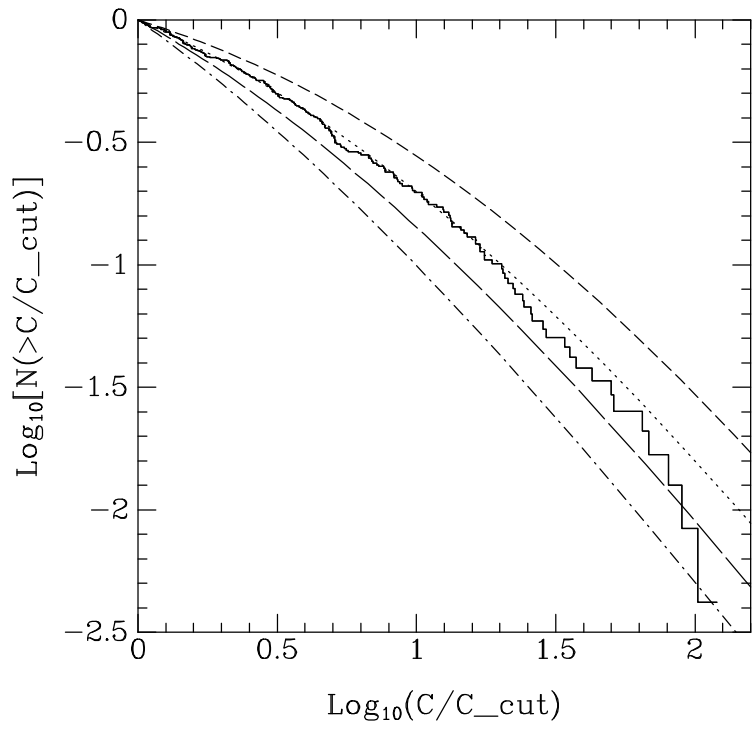

Fig. 1

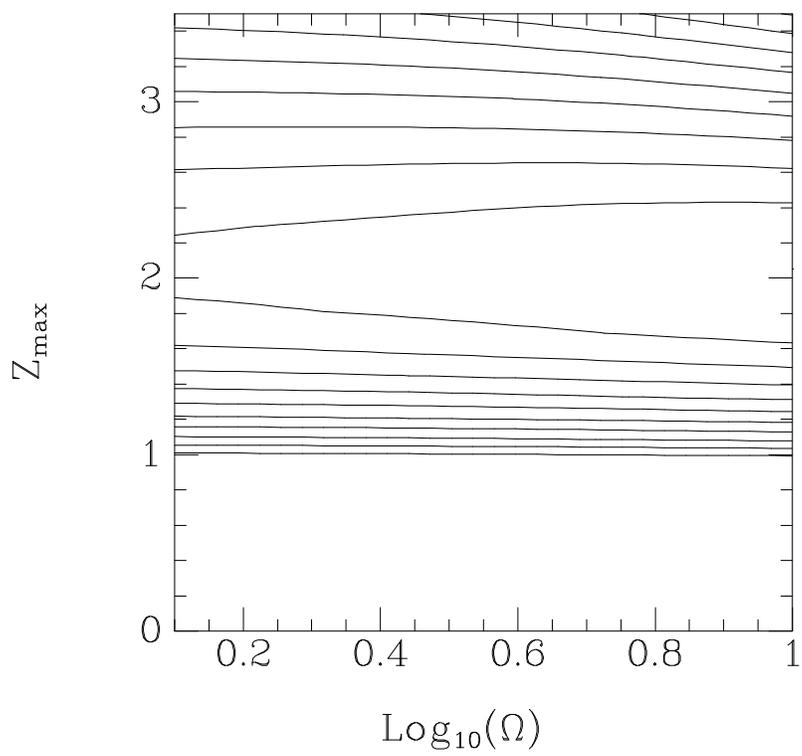

Fig. 2

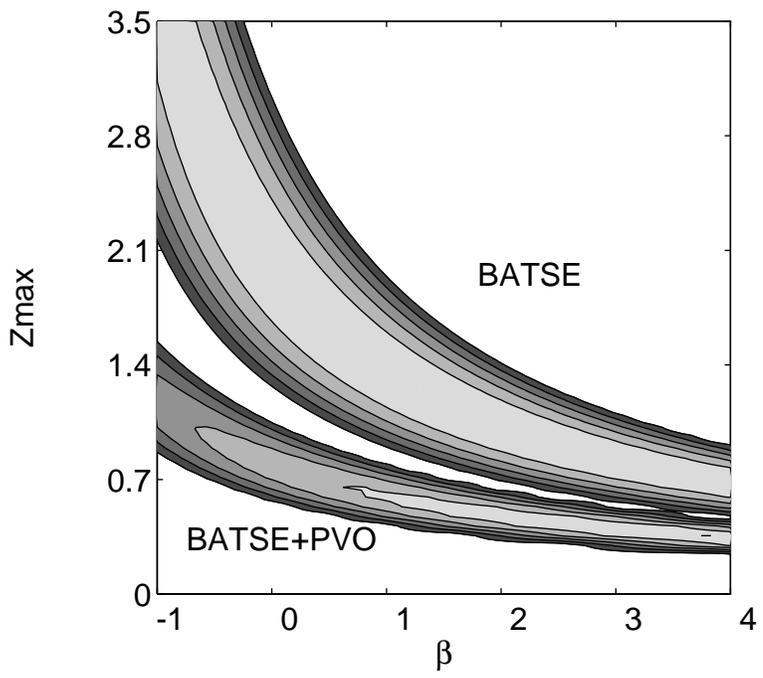

Fig. 3

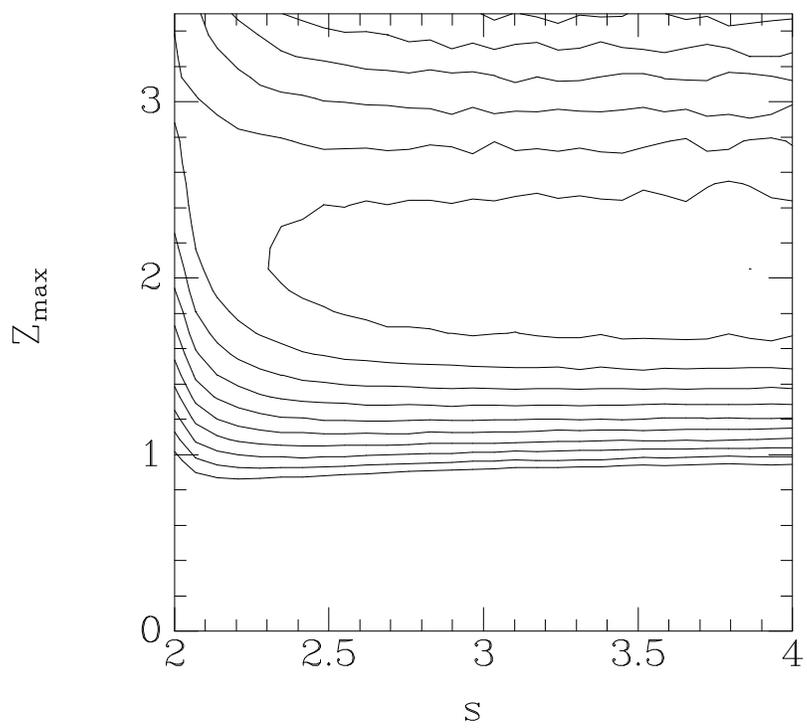

Fig. 4